\documentclass[journal]{IEEEtran}

\usepackage{cite}

\usepackage{graphicx}
\usepackage{amsmath, amsthm, amssymb}
\usepackage{float}
\usepackage{indentfirst}
\usepackage[utf8]{inputenc}
\usepackage[dvipsnames]{xcolor}

\graphicspath{{imagens/}}

\begin{document}

\title{Optimized Quantization in Distributed Graph Signal Filtering}

\author{Isabela~Cunha~Maia~Nobre
        and~Pascal~Frossard

\thanks{Isabela Nobre and Pascal Frossard are with the Signal Processing Laboratory (LTS4), École Polytechnique Fédérale de Lausanne (EPFL), Switzerland. (e-mails: isabela.nobre@epfl.ch, pascal.frossard@epfl.ch).}
\thanks{Isabela Nobre is partially supported by CAPES (grant number 88881.174577/2018-01).}}

\maketitle

\begin{abstract}
Distributed graph signal processing algorithms require the network nodes to communicate by exchanging messages in order to achieve a common objective. These messages have a finite precision in realistic networks, which may necessitate to implement message quantization. Quantization, in turn, may generate distortion and performance penalty  in the distributed processing tasks. This paper proposes a novel method for distributed graph filtering that minimizes the error due to message quantization without compromising the communication costs. It first bounds the exchanged messages and then allocates a limited bit budget in an optimized way to the different messages and network nodes. In particular, our novel quantization algorithm adapts to both the network topology and the message importance in a distributed processing task. Our results show that the proposed method is effective in minimizing the error due to quantization and that it permits to outperform baseline distributed algorithms when the bit budget is limited.  They further show that errors produced in nodes with high eccentricity or in the first steps of the distributed algorithm contribute more to the global error. Also, sparse and irregular graphs require more irregular bit distribution. Our method provides one of the first quantization solutions for distributed graph processing, which is able to adapt to the target task, the graph properties and the communication constraints. 
\end{abstract}


\section{Introduction}
\label{sec:intro}

Many different networks, for instance wireless sensor networks, transportation networks, neural networks and social networks can be modeled as graphs where nodes support signal values or features. The field of signal processing on graphs  has been providing many powerful tools to process such signals in diverse applications, such as compression, denoising or reconstruction of sensor data \cite{ortega2017graph,6494675}. 

Numerous applications require that the signal defined on the network is however processed distributively. Decentralized methods of graph signal processing recently emerged in order to scale to large networks as a way to deal with big data applications, privacy concerns and also bandwidth/energy constraints \cite{DBLP:journals/corr/abs-1111-5239, sandryhaila2014finite,safavi2015revisiting}. They can also be necessary when centralized topologies are not viable or suffer from a bottleneck on the central processor, with nodes that are too far to reach the central processor. Another advantage of distributed algorithms compared to centralized ones is that they can add robustness to the network in case of node failures. 

In practice, most nodes in distributed systems have limited computing power and are constrained in the amount of information they can communicate; this makes the design of efficient distributed algorithms essential. In order to enable distributed graph signal processing, linear graph signals operators can for example be approximated by shifted Chebyshev polynomials \cite{DBLP:journals/corr/abs-1111-5239}, becoming more amenable to distributive computing for applications such as smoothing, denoising, inverse filtering and semi-supervised learning. There have been many other studies on distributed processing for graph signals or networked data \cite{isufi2017autoregressive,segarra2017optimal,sandryhaila2014finite,safavi2015revisiting,shi2015infinite}, but only few deal with the fact that, in real case scenarios, the network is subject to communication constraints that limit the precision of the messages exchanged by distributed algorithms.

In this paper, we propose an adaptive quantization scheme for distributed graph signal processing tasks. Quantization approximates a continuous range of values by a set of discrete values and helps decreasing the communication costs. At the same time, it introduces errors that degrade the performance of distributed processing methods.   We build on our previous work \cite{8682784} and focus on the design of a quantization scheme that minimizes the quantization error in graph signal processing tasks, by bounding the transmitted messages and by optimizing the bit allocation. We first propose a distributed processing algorithm where the messages exchanged by network nodes are bounded. We then model the error due to quantization, as a function that depends on the network topology and on the characteristics of the distributed processing task. We further cast an optimization problem to efficiently distribute bit to messages and nodes, so that the total error is minimized under communication cost constraints. We finally solve the resulting bit allocation problem with the Karush-Kuhn-Tucker (KKT) conditions. The performance of our quantizer is evaluated and  compared to the performance of an uniform quantizer. The results show that the bit allocation optimization 
improves the performance in terms of final error compared to
a uniform distribution of the bit budget across messages. They also show that a more regular graph leads towards more uniform bit distribution in the optimal allocation, which confirms the proper adaptation of our solution to the network properties. Also, since the errors propagate through the successive steps of the distributed signal processing operations, we also confirm that it is efficient to allocate a large share of the bit budget in the first steps of the iterative distributed processing algorithm.

Some works have considered different aspects of quantization in graph signal processing. The works in \cite{sandryhaila2013discrete} and \cite{liu2017filter} briefly studied the effect of quantization in a linear prediction filter applied to graph signals. However the main objective of those works was to design graph filters, without considering specifically the quantization effects in the design itself. On the other hand, the work in \cite{chamon2017finite} studied the effect of quantization in the representation of graph signals, mitigating the 
numerical effects caused by the finite-precision machines that centrally realize the filtering process. The authors specifically designed graph filters that are
robust to finite precision effects.  The above works had a centralized graph filtering approach.  There are also some works that focused on distributed optimization with quantization constraints \cite{zhu2016quantized,zhu2016quantized2,li2017event, li2017distributed,thanou2013distributed}, but these algorithms have been developed mainly using consensus protocols, and we are  interested in solving more general processing tasks in this paper, not merely average computation. Closer to our framework, the work \cite{7447171} derived the quantization error in distributed computations of graph signal operators and then proposed an algorithm that learns graph dictionaries to sparsely approximate graph signals while staying robust to the quantization noise. The work however does not focus on improving the quantization but rather on the design of robust graph filters.  To the best of our knowledge,our work the first one that optimizes the quantization scheme for general distributed filtering tasks.

The rest of this paper is organized as follows. In Section 2, we model the graph filter implementation in a distributed way and derive the quantization error. In Section 3, we present our new bounding scheme for processing graph signals distributively, and in Section 4 we describe our optimal bit allocation algorithm. Finally, the results and analysis of the performance of our new framework are shown in Section 5.

\section{Distributed Graph Signal Processing} \label{SecZ}

\subsection{Background}
We first consider that the network can be represented by a weighted, undirected graph  $ \mathcal{G}  = (\mathcal{V}, \mathcal{E}, W) $, where $ \mathcal{V} $ represents a set of vertices, $ \mathcal{E} $ represents a set of edges and $W$ is the weight matrix whose entries $W[i,j]$ typically depend on the distance between nodes $i$ and $j$. The number of nodes in the graph is $ N = |\mathcal{V}|  $. We define $D$ as the diagonal degree matrix whose elements are the sum of each row of $W$ respectively. The normalized graph Laplacian operator is frequently used in graph signal processing. It is given by $ \mathcal{L} = I - D^{-1/2}WD^{-1/2} $, which is a real symmetric positive semi-definite matrix. We further denote as  $\{\lambda_n\}_{n=0..N-1}$  the set of eigenvalues of $ \mathcal{L} $, which we order as $ \left\lbrace 0=\lambda_0 < \lambda_1\leq \lambda_2 \leq ..\leq \lambda_{N-1}\leq 2\right\rbrace  $, and its eigenvectors as $ \left\lbrace \mathcal{X}_0,\mathcal{X}_1,...,\mathcal{X}_{N-1}\right\rbrace  $.  Finally, we denote by $\Lambda$ the diagonal matrix with the eigenvalues $\{\lambda_n\}_{n=0..N-1}$ on its diagonal and $\mathcal{X}$ the matrix whose columns are the eigenvectors of $ \mathcal{L} $.

A graph signal is a function $f:\mathcal{V}\longrightarrow \mathbb{R}$ defined on the vertices of the graph, which is represented by a vector $\textbf{f} \in \mathbb{R}^N$. The graph Fourier transform of $f $ can be defined as an expansion of the function in terms of the eigenvectors of $\mathcal{L}$, that is $\hat{f}(\lambda_n) = \sum_{i=1}^{N}f(i)\mathcal{X}_n(i)$ \cite{6494675}. Given $ g(\lambda)  $ the transfer function of a graph filter,  we can process the signal $f_{in}$ by computing  $ f_{out} = g(\mathcal{L})f_{in} $, where 
\begin{equation} \label{eqA}
g(\mathcal{L}) :=  \mathcal{X} \begin{bmatrix}
g(\lambda_0)     &  & 0 \\
&  &  \\
0   &   & g(\lambda_{N-1}) 
\end{bmatrix}   \mathcal{X}^T. 
\end{equation}

\subsection{Distributed Filtering}

As many graph signal processing tasks can be represented as filtering operations, we focus on operators of the form of (\ref{eqA}) in the rest of the paper. If the signal processing operator $ g(\mathcal{L} ) =   \sum\limits_{k=0}^{K}\alpha_k \mathcal{L}^k $ is a polynomial function of order $K$ of the Laplacian (or if it can be approximated by a polynomial, which is usual in classical settings), with $ \{\alpha_k\}_{k=0..K} $ as the polynomial coefficients, it is amenable to distributed implementation  \cite{DBLP:journals/corr/abs-1111-5239}. The corresponding graph filter
\begin{equation} \label{eqbbb}
f_{out} = \sum\limits_{k=0}^{K}\alpha_k \mathcal{L}^kf_{in} 
\end{equation} 
can be implemented iteratively by local processing at each node, and message exchanges between nodes. That is, each node computes the local value of the function $f_{out}$ by exchanging messages with neighbors in $K$ interactions. Along this process, all nodes together participate in computing the full function $f_{out}$. 

The distributed implementation requires the computation of the different powers of the Laplacian matrix that appear in (\ref{eqbbb}). We firstly define  $z_k=\mathcal{L}^kf_{in}$ and begin with
$z_0 = f_{in}$. The node $n$ sends its value $z_0[n]$ to its one-hop neighbors in the graph, and all the other nodes do the same. After all values of $z_0$ are exchanged in the network, all nodes update their local status with the relation $z_1=\mathcal{L} z_0$. To that end, each node $n$ will only calculate its value $z_1[n]$ by doing $\mathcal{L}_n^Tz_0$, where $\mathcal{L}_n$ is the line $n$ of $\mathcal{L}$; $z_0$ is filled with the values of the messages exchanged from the neighbors of node $n$. Since  $\mathcal{L}_n$ is zero for the nodes that are not neighbors of $n$, the node $n$ does not need the values of $z_0$ at these nodes to calculate $z_1[n]$. 

The messages with the values of $z_1$ are then exchanged between neighbor nodes in the same way as for the values of $z_0$. Then $z_2=\mathcal{L} z_1$ is obtained in a similar fashion as $z_1$. This procedure repeats until $K$ iterations, which permits to compute the full function $f_{out}$ in Eq. (\ref{eqbbb}). Specifically, after knowing $\{z_0[n],z_1[n],.. z_K[n] \} $, the node $n$ computes the value of the filtered signal in its own node using the relation \cite{7447171} 
\begin{equation}\label{eqB}
f_{out}[n]=\left( g(\mathcal{L})f_{in}\right) [n]  = \sum\limits_{k=0}^{K}\alpha_k z_k[n].
\end{equation}
More details are given in \cite{DBLP:journals/corr/abs-1111-5239}. 

\subsection{Quantization error}

The above computation permits to perfectly compute the response of the graph filter in distributed settings. However, in realistic cases, the messages exchanged by the network nodes have a limited precision. They are typically quantized before transmission and this modifies the outcome of the distributed filtering process. 

The quantized message at node $n$ at step $k$ can be written as $
\tilde{z}_k[n] = z_k[n]+\epsilon_k[n]$,
where  $ \epsilon_k[n] $ is the quantization error for a message at iteration step $k$ at node $n$. After its transmission, the distributed update equation at node $n$ becomes $
z_{k+1} = \mathcal{L} \tilde{z}_k[n] = \mathcal{L}(z_k[n]+\epsilon_k[n] )$ and integrates a quantization error term.

We define by
\begin{equation}\label{epislon}
\epsilon_k = \begin{bmatrix}
\epsilon_k[0] \\
\vdots \\
\epsilon_k[N]
\end{bmatrix}
\end{equation}
the vector containing error values for all nodes at step k, and  $\epsilon = [\epsilon_0^T,...,\epsilon_{K-1}^T]^T$ the vector containing all errors at all iteration steps and nodes. 

By taking into account the  quantization errors that accumulate through all iterations of the distributed processing task, the filtered signal can finally be written as 
\begin{equation}\label{eqblabla}
g(\mathcal{L})f_{in}  = \sum\limits_{k=0}^{K}\alpha_k \mathcal{L}^kz_0 +  \sum\limits_{k=0}^{K-1}\left[ \sum\limits_{j=1}^{K-k}\alpha_{k+j}\mathcal{L}^j\right] \epsilon_k,
\end{equation}
as opposed to $f_{out}$ in Eq. (\ref{eqbbb}) for the perfect settings. More details are available in \cite{7447171}.

\section{Quantized Distributed Filtering with bounded messages}

\subsection{Quantization error analysis}

We first analyze more deeply the impact of quantization. Since the first term of the above expression (\ref{eqblabla}), $ \sum\limits_{k=0}^{K}\alpha_k \mathcal{L}^kz_0 $, is the filtered signal in a setting without quantization (as in Eq (\ref{eqB})), we can define the second term, 
\begin{equation} \label{eqF}
Q = \sum\limits_{k=0}^{K-1}\left[ \sum\limits_{j=1}^{K-k}\alpha_{k+j}\mathcal{L}^j\right] \epsilon_k,
\end{equation}
as the total error caused by the accumulated effects of the quantization errors.

\label{SecA}

We can further make the following observation on the evolution of the quantization error with the iterations of the distributed processing algorithm. At step $k$, the maximum value (in an absolute sense) of the messages to be transmitted is $ || \mathcal{L}^kf_{in}||_\infty $.
Considering that, for $p>r>0$,  we have $ \parallel v \parallel_r \:\:\geq \:\:\parallel v \parallel_p  $, for any $v$ pertaining to the vector space where these norms are defined, we can write
\begin{equation}
\parallel \mathcal{L}^kf_{in} \parallel_\infty \:\: \leq \:\:\parallel \mathcal{L}^kf_{in} \parallel_2\:\: \leq \:\:\parallel \mathcal{L}^k \parallel_2\cdot \parallel f_{in} \parallel_2,
\end{equation}
where $ \parallel \mathcal{L}^k \parallel_2 $ is a matrix norm induced from the 2-norm for vectors, which can be computed by  $
\parallel \mathcal{L}^k \parallel_2 = \sigma_{max}(\mathcal{L}^k ) $,
where $ \sigma_{max}(\mathcal{L}^k ) $ represents the largest singular value of matrix $ \mathcal{L}^k  $ \cite{meyer2000matrix}. It corresponds to the square root of the largest eigenvalue of the positive-semidefinite matrix $ (\mathcal{L}^k )^T(\mathcal{L}^k ) $. Since the Laplacian matrix is diagonalizable and symmetric, we can write 
$ (\mathcal{L}^k )^T = (\mathcal{L}^k ) $, and since 
\begin{equation}
\mathcal{L} = \mathcal{X}	\Lambda \mathcal{X}^T,
\end{equation}
we have
\begin{equation}
\mathcal{L}^{2k} = \mathcal{X}	\Lambda^{2k} \mathcal{X}^T,
\end{equation}
which means that the eigenvalues of $ \mathcal{L}^{2k} $  are the same as the eigenvalues of $ \mathcal{L} $ to a power of $ 2k $. Hence, since
\begin{equation}
\parallel \mathcal{L}^k \parallel_2 = \sigma_{max}(\mathcal{L}^k ) = \sqrt{\lambda_{max}(\mathcal{L}^{2k})} =  \sqrt{\lambda_{N-1}^{2k}},
\end{equation}
and considering that all eigenvalues of the Laplacian are real and positive values, we finally have
\begin{equation}\label{eqL}
\parallel \mathcal{L}^kf_{in} \parallel_\infty \:\: \leq  \:\: \lambda_{N-1}^k \cdot \parallel f_{in} \parallel_2.
\end{equation}

This means that, as $k$ increases, the transmitted messages can increase their ranges proportionally to the eigenvalues of $\mathcal{L}$, as shown in Eq. (\ref{eqL}). This also means that, at a high value of $k$, the value of the respective error $\epsilon_k$ will be very high, hence increasing the value of the total error. We show below how to bound these transmitted messages such that the quantization errors will remain bounded as well.

\subsection{Distributed processing with bounded messages}

Based on the previous observations, we propose a modification of the classical distributed processing algorithm that permits to bound the range of the exchanged messages. Instead of using the operator given by the normalized Laplacian $ \mathcal{L} $ at every step of the distributed algorithm, we use \label{sec:bound}
\begin{equation}
\dot{\mathcal{L}}=\mathcal{L}-I. 
\end{equation}
Hence, the eigenvalues of $ \dot{\mathcal{L}} $ will be bounded in $[-1,1]$, instead of $[0,2]$, with $ \mathcal{L} $. Therefore the values of the messages being transmitted at step $k$ will surely not surpass the range of the original signal $z_0=f_{in}$, as shown in Eq. (\ref{eqL}).

In order to integrate the modified Laplacian operator $ \dot{\mathcal{L}} $, we modify the distributed filtering algorithm of the previous section as follows. We start with a scenario without quantization. Firstly we set $z_0 = f_{in}$, which is then exchanged with the neighbor nodes, as before. Now, instead of multiplying the received values by $ \mathcal{L} $, the nodes rather compute $\dot{z}_1 = \dot{\mathcal{L}}z_0 $. The value $\dot{z}_1 $ is then exchanged with neighbor nodes, in a similar way as the algorithm described in the previous section. The algorithm then proceeds iteratively in the same way. 

In an ideal scenario (without quantization), we have 
$\dot{z}_k = \dot{\mathcal{L}}^kz_0 = (\mathcal{L}-I)^kz_0$. We observe that, at each step, we can perfectly recover  $z_k = \mathcal{L}^kz_0$ from $\dot{z}_k $. Since the identity matrix commutes with all matrices, $\mathcal{L}$ and $I$ also commute. Hence, we can use the Binomial formula and derive
\begin{equation}\label{eqN3}
\dot{z}_k = \left( \sum_{i=0}^{k}\binom{k}{i}(-1)^{k-i}\mathcal{L}^i\right) z_0  =  \sum_{i=0}^{k}\binom{k}{i}(-1)^{k-i} z_i,
\end{equation}
or equivalently,
\begin{equation}\label{eqN}
z_k = \dot{z}_k-\sum_{i=0}^{k-1}\binom{k}{i}(-1)^{k-i}z_i  =  \sum_{i=0}^{k}\binom{k}{i}\dot{z}_i.
\end{equation}

Therefore, with Eq. (\ref{eqN}), we can build a distributed algorithm where the values of $ \dot{z}_k $ are exchanged between neighbor nodes, but only the values $ z_k $ corresponding to the iterative solutions of the original distributed filtering algorithm need to be stored. This is useful if the sensors have memory constraints. Note that the term $\sum_{i=0}^{k}\binom{k}{i}\dot{z}_i$ is purely combinatorial, that is, it does not depend on any specific component such as the network, data, iteration or task, etc. 

\subsection{Quantization error with bounded messages}

In a scenario with quantization, the distributed filtering algorithm with the modified Laplacian operator $ \dot{\mathcal{L}} $ is modified as follows. After we set $z_0 = f_{in}$, we quantize it as $\tilde{z}_0 = z_0 +\epsilon_0$. The values $\tilde{z}_0$ are then exchanged with the neighbor nodes, as before. Now, instead of multiplying the received values by $ \mathcal{L} $ as in the original algorithm, the nodes rather compute $\dot{z}_1 = \dot{\mathcal{L}}\tilde{z}_0 $ in the bounded scheme, the resulting value is then quantized as $\tilde{\dot{z}}_1 = \dot{z}_1 +\epsilon_1$. The quantized value $\tilde{\dot{z}}_1 $ is eventually exchanged with neighbor nodes, in a similar way as the algorithm described above.

To recover $z_k$ from $\dot{z}_k$ we use the same process as in Eq. (\ref{eqN}). However, the recovery is not perfect anymore due to quantization. The quantization error now accumulates through iterations and the value of $ \dot{z}_k $ becomes 
\begin{equation}\label{eqO}
\dot{z}_k=(\mathcal{L}-I)^kz_0+\sum_{l=0}^{k-1}(\mathcal{L}-I)^{k-l}\epsilon_l, \;\;\;\;\;\;\;\;\; \text{for    }   k>0,
\end{equation} 
while it is given by Eq. (\ref{eqN3}) in the ideal settings. More specifically, we compute below the quantization error in receiving the output of the filtering process. First, Eq. (\ref{eqN}) is equivalent to 
\begin{equation}\label{eqN2}
z_k = z_0 +  \sum_{i=1}^{k}\binom{k}{i}\dot{z}_i.
\end{equation}
If we replace $ \dot{z}_i $ in (\ref{eqN2})  with the expression in  (\ref{eqO}), we obtain
\begin{equation}\label{eqN4}
z_k = z_0 +  \sum_{i=1}^{k}\binom{k}{i}\left[ (\mathcal{L}-I)^iz_0+\sum_{l=0}^{i-1}(\mathcal{L}-I)^{i-l}\epsilon_l\right].
\end{equation}
Notice that, if we write $\mathcal{L}^k$ as $[(\mathcal{L}-I)+I]^k$ and use the Binomial formula, we obtain
\begin{equation}
\mathcal{L}^k = \sum_{i=0}^{k}\binom{k}{i} (\mathcal{L}-I)^i.
\end{equation}
Thus, (\ref{eqN4}) can be written as
\begin{equation}
z_k=\mathcal{L}^kz_0+\sum_{i=1}^{k}\binom{k}{i}\sum_{l=0}^{i-1}(\mathcal{L}-I)^{i-l}\epsilon_l, \;\;\;\;\;\;\;\;\; \text{for    }   k>0.
\end{equation} 
The distributed filtering of the graph signal $f_{in}$ in the quantization regime can then be written as
\begin{equation}\label{eqaa}
\dot{f}_{out} = g(\mathcal{L})f_{in}  = \sum\limits_{k=0}^{K}\alpha_k \mathcal{L}^kz_0 +\sum\limits_{k=1}^{K}\alpha_k \sum_{i=1}^{k}\binom{k}{i}\sum_{l=0}^{i-1}(\mathcal{L}-I)^{i-l}\epsilon_l,
\end{equation}
from which we derive the total error, which corresponds to the second term in Eq. (\ref{eqaa}), 
\begin{equation} 
\dot{Q}  = \sum\limits_{k=1}^{K}\alpha_k \sum_{i=1}^{k}\binom{k}{i}\sum_{l=0}^{i-1}(\mathcal{L}-I)^{i-l}\epsilon_l.
\end{equation}

It can be rewritten as
\begin{equation} \label{eq42}
\dot{Q}  = \sum\limits_{k=0}^{K-1}\left[ \sum_{i=1}^{K-k}\alpha_{k+1} \sum_{j=1}^{i}\binom{k+i}{k+j}(\mathcal{L}-I)^{j}\right]\epsilon_k.
\end{equation}

For the sake of clarity, we now write 
\begin{equation}\label{eqs}
F_k[n] =	\left( H_k^TH_k \right) [n,n],
\end{equation}
with 
\begin{equation}\label{eqr}
H_k =  \sum_{i=1}^{K-k}\alpha_{k+1} \sum_{j=1}^{i}\binom{k+i}{k+j}(\mathcal{L}-I)^{j}.
\end{equation}

If we combine (\ref{eqr}) and (\ref{eqs}), we get
\begin{multline}\label{eqt}
F_k[n] = \sum_{i=1}^{K-k}\alpha_{k+i} \sum_{j=1}^{i}\binom{k+i}{k+j}\sum_{p=1}^{K-k}\alpha_{k+p}\sum_{q=1}^{p}\binom{k+p}{k+q}\cdot \\
(\mathcal{L}-I)^{j+q}[n,n].
\end{multline}

Hence,
\begin{equation} 
\dot{Q}  = \sum\limits_{k=0}^{K-1}H_k\epsilon_k,
\end{equation}
and we can finally calculate the mean squared error
\begin{equation} \label{equai}
\parallel \dot{Q}  \parallel ^2  = \sum\limits_{k=0}^{K-1} \sum\limits_{l=0}^{K-1}\epsilon_k^TH_k^TH_l\epsilon_l,
\end{equation}
and its expected value 
\begin{equation} \label{coisa}
E_{\epsilon}      \left[        \parallel \dot{Q}  \parallel ^2    \right]   = \sum\limits_{k=0}^{K-1} \sum\limits_{l=0}^{K-1}E_{\epsilon}[\epsilon_k^TH_k^TH_l\epsilon_l].
\end{equation}

\section{Optimized bit allocation}
\subsection{Rate-distortion model}

In this Section, we now seek to minimize the expected value of the square of the total error in Eq. (\ref{equai}) so that the impact of communication constraints on the distributed computations is minimum. 

The quantization error is a deterministic function of the signal  $f_{in}$. Thus, given each $f_{in}$, the quantization error is deterministic. But we would like to have an optimal quantization scheme for all input signals. In order to obtain that, we model the input signal at each node as random variables and by consequence, we model the error signal as a random variable as well  \cite{492748}. Thus, our aim is to minimize the expected value of the total mean square error for these random variables.

We define $x[n,k]$ as the number of bits used to represent the message sent from node $n$ at step $k$. We consider the high rate regime and use uniform quantizers for each message. In this case, the error is directly related to the number of bits used for each message. The quantization step size is then determined by the ratio of the total quantization range over the number of quantization intervals, that is
\begin{equation} \label{equ}
\Delta[n,k] =  \frac{2 \cdot \parallel f_{in}\parallel_\infty}{2^{x[n,k]}},
\end{equation} 
the quantization error is $2 \cdot \parallel f_{in}\parallel_\infty $ for al $k$ since we filter the signal with the modified Laplacian $\dot{\mathcal{L}}$ to bound the messages.

It had been shown \cite{492748} that even though the quantization noise and input signal are deterministically related, when the quantization step size is sufficiently small (that is, in high rate conditions), the quantization noise is uncorrelated with the quantized signal. Furthermore, the probability density function of the quantization noise will be uniform with zero mean and variance $\frac{\Delta[n,k]^2}{12}$. This is called the White-Noise Quantization Error Model  \cite{1166538}.

It follows that the expected value of the square of the
quantization error on the message transmitted by node $n$ at step $k$ is equal to its variance (since it has zero mean), and if we apply (\ref{equ}) into the variance expression of the White-Noise Quantization Error Model, we obtain
\begin{equation}\label{gg}
E_{\epsilon}[\epsilon_k[n]^2] = \frac{  \parallel f_{in}\parallel_\infty^2}{3}\cdot2^{-2x[n,k]}.
\end{equation} 
With the White-Noise Quantization Error Model, we can assume that  $\epsilon_k[n]$ and $\epsilon_p[m]$ are statistically independent for $k\neq p$ or $n\neq m$. Hence the expected value of the crossed terms in Eq. (\ref{coisa}) is zero and we finally obtain the expected value of the total mean squared error as  
\begin{equation} 
E_{\epsilon}[\parallel \dot{Q}  \parallel ^2]  = \sum\limits_{k=0}^{K-1} \sum\limits_{n=0}^{N-1}	\left( H_k^TH_k \right) [n,n] E_{\epsilon}[\epsilon_k[n]^2],
\end{equation}
which is equivalent to
\begin{equation}\label{seodfdsfds}
E_{\epsilon}[\parallel \dot{Q}  \parallel ^2] = \sum\limits_{k=0}^{K-1} \sum\limits_{n=0}^{N-1} F_k[n]  E_{\epsilon}[\epsilon_k[n]^2], 
\end{equation} 
with (\ref{gg}), the MSE finally reads
\begin{equation}\label{bumbum}
E_{\epsilon}[\parallel \dot{Q}  \parallel ^2] =  \frac{\parallel f_{in}\parallel_\infty^2}{3}\sum\limits_{k=0}^{K-1} \sum\limits_{n=0}^{N-1} F_k[n]  2^{-2x[n,k]}.
\end{equation} 

\subsection{Optimal Allocation}

Our objective is now to minimize the total quantization error given the constraint bit budget in the network. To that end, it is necessary to find the values of $x[k,n]$ for every combination of $k$ and $n$ that obey the budget constraint and minimize $ E_{\epsilon}[\parallel \dot{Q}  \parallel ^2] $  of Eq. (\ref{bumbum}). This optimization can be described by the following problem:

\begin{equation}\label{oti}
\boxed{\begin{aligned}
	& \underset{x}{\text{minimize}}
	& & E_{\epsilon}[\parallel \dot{Q}  \parallel ^2]  \\
	& \text{subject to}
	& & \sum\limits_{k=0}^{K-1} \sum\limits_{n=0}^{N-1}x[n,k]d[n]\leq B
	\end{aligned}}
\end{equation}

Here, $ d[n]  $ is the degree of node $n$, which drives the transmission costs, and $B$ is the total bit budget constraint over the whole network. The term $ x[n,k]d[n] $ represents the total number of bits used by node $n$ at step $k$ to send the respective message for all $d[n]$ neighbors. When we sum the $ x[n,k]d[n] $ term for every $k$ and $n$, we obtain the total number of bits used to process the signal, which is our constraint in the optimization problem. Since the objective and the constraint functions are both continuously differentiable and convex on the values of $x[n,k]$, we can use the Karush-Kuhn-Tucker (KKT) conditions and they will be sufficient for finding the optimal solution. Also, we observe that the problem satisfies the KKT regularity conditions since the constraint function is affine \cite{boyd2004convex}. 

The KKT stationarity condition states that  the solution of the optimization problem (\ref{oti}) will be the values of $x[n,k]$ that satisfy the equation
\begin{equation}\label{zz}
\frac{\partial}{\partial x[n,k]} \left[ (E_{\epsilon}[\parallel \dot{Q}  \parallel ^2]) +\mu  (\sum\limits_{k=0}^{K-1} \sum\limits_{n=0}^{N-1}x[n,k]d[n]-B)\right] = 0,
\end{equation}
where $ \frac{\partial}{\partial x[n,k]} $ means the partial derivative with respect to $  x[n,k] $, and $ \mu  $ is the KKT multiplier. The solution to Eq. (\ref{zz}) is given by
\begin{equation} \label{eqw}
x[n,k] = -\frac{1}{2\ln(2)}\ln\left( \frac{3\:\mu\: d[n]}{2\parallel f_{in}\parallel_\infty^2\ln(2)F_k[n]}\right).
\end{equation}

The complementary slackness condition states that
\begin{equation}\label{eqa}
\mu (\sum\limits_{k=0}^{K-1} \sum\limits_{n=0}^{N-1}x[n,k]d[n]-B)= 0,
\end{equation}
but $\mu$ cannot be equal to zero, otherwise the expression in (\ref{eqw}) would present a $\ln(0)$ term, which is undefined. This leads to the second term of Eq. (\ref{eqa}) being equal to zero instead, and applying (\ref{eqw}) into (\ref{eqa}) and solving for $\mu$, it results into
\begin{equation}
\mu=\exp\left( {-\frac { B\ln(4)+ \sum \limits_{n=0}^{N
			-1} d[n] \sum \limits_{k=0}^{K-1}\ln  \left(    {\frac {3d[n]}{2\parallel f_{in}\parallel_\infty^2
				\ln	2\cdot F_k[n]}} \right)  }{K\sum\limits _{n=0}^{N-1}d[n]} }\right) ,
\end{equation}
which is always non-negative for being an exponential function, thus guaranteeing that the KKT dual feasibility condition is satisfied. And we can finally replace in (\ref{eqw}) to obtain the optimized number of bits for each message. 

In practice, a conventional approach \cite{1055533} is to further round the non-integer values in Eq. (\ref{eqw}) to become integers. Then, if some of the values of $x[n,k]$ obtained in (\ref{eqw}) are negative or zero, they are replaced by 1, to guarantee that there is a minimal communication between neighboring nodes in every step to keep the whole system synchronized. 

We can analyze the optimal bit allocation solution as follows. We observe that the number of bits $x[n,k]$ in Eq. (\ref{eqw}) depends on $d[n]$ and $F_k[n]$. As the degree $d[n]$ grows, the communication cost in node $n$ grows as well, since it shares its messages with more neighbors. In order to satisfy the budget constraint, $x[n,k]$ has thus to be smaller in a node where $d[n]$ is larger. On the other hand, the factor $F_k[n]$ is related to the topology and to the filter coefficients. From Eq. (\ref{seodfdsfds}), we can see that it weights the contribution of each individual error $\epsilon_k[n]$ in the global error. If we have a big $F_k[n]$, it means that the relative contribution of $\epsilon_k[n]$ becomes big, so that its contribution needs to be reduced by increasing  $x[n,k]$. Also, for the same node $n$, there are more error terms in the global error computation (in Eq. (\ref{eqt})) when $ k $ is low, which means that $F_k[n]$ becomes higher in this case. It further means that the relative contribution of the error terms $\epsilon_k$ in the first iterations (small $k$) of the distributed processing algorithm is higher compared to the error in the later iterations. This is in agreement with the fact that the first error terms lead to higher propagation effects. These observations will be further confirmed in the upcoming result section.

\section{Results}

\subsection{Performance of the proposed scheme}

\subsubsection{Performance of the optimal quantizer} \label{sub_perf_opt_quant}
The performance of our quantizer is now evaluated in details. First, we create a network with $N=50$ nodes that are uniformly placed at random in a unit square. The weight matrix for the network edges is generated based on a thresholded Gaussian kernel function that takes into account the physical distance between nodes. The edges weights are given by $W_{ij}=e^{-l_{ij}^2/\theta}$ if the distance $l_{ij}$ between vertices $i$ and $j$ is less or equal to $\kappa$, and zero otherwise. We fix $\kappa=0.2$. 

A graph signal is defined as $ f_{in}[n]= a[n]^2+b[n]^2-1 $, where $a[n]$ and $b[n]$ are the coordinates of node $n$,  
and a random noise with zero-mean normal distribution is added to it. We consider denoising as the distributed graph signal processing task, via a low-pass filter 
\begin{equation}\label{filtro}
g(\lambda) = \frac{\tau}{\tau+5\lambda} 
\end{equation}
with $\tau=3$. In order to  implement it distributively, a Chebyshev polynomial approximation of order $K$ is performed, and its polynomial coefficients $ \{\alpha_k\}_{k=0..K} $ are determined as in \cite{DBLP:journals/corr/abs-1111-5239}. 

\begin{figure}[htb]
	\centering
	\centerline{\includegraphics[width=\linewidth]{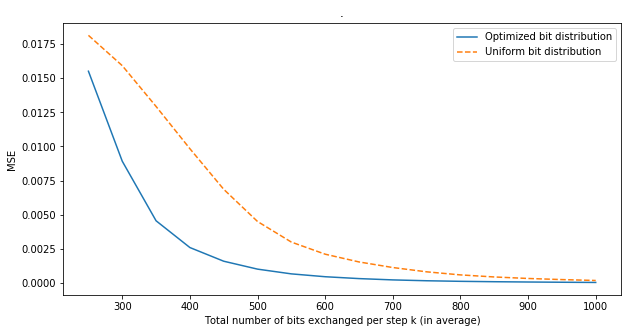}}
	\caption{Average MSE vs average number of bits for uniform, and optimized bit allocation for $K=9$ and $\theta=2$ ($\kappa=0.2$, $ f_{in}[n]= a[n]^2+b[n]^2-1 $,  $ g(\lambda) = \frac{3}{3+5\lambda}  $). }\medskip
	\label{fig:resultadobounded}
\end{figure}

The distributed graph signal processing task is first performed without quantization. Then, the processing is performed with uniform quantization that is used as baseline solution. In this case, the number of bits used to represent the transmitted messages is the same for every node $n$ and iteration step $k$. Finally, another processing is performed, using the optimization scheme described in this paper.  In all three cases, the bounding scheme from Section \ref{sec:bound} is used. We calculate the MSE between the output of the quantized and the unquantized schemes for both uniform and optimized quantization. We repeat the entire experiment 1000 times for different networks and compute average performance shown in Figure \ref{fig:resultadobounded}.

\begin{figure}[htb]
	\centering
	\centerline{\includegraphics[width=\linewidth]{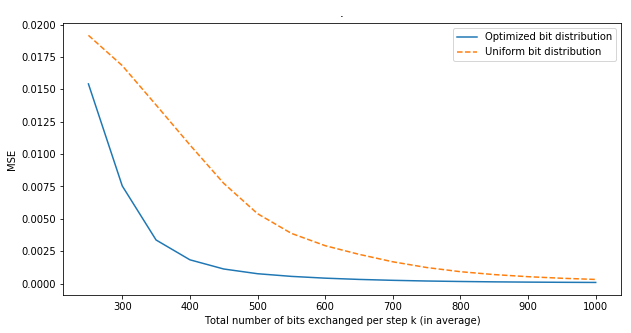}}
	\caption{Average MSE vs average number of bits for uniform and optimized bit allocation for $K=9$ and $\theta=0.001$ ($\kappa=0.2$, $ f_{in}[n]= a[n]^2+b[n]^2-1 $,  $ g(\lambda) = \frac{3}{3+5\lambda}    $).}
	\medskip
	\label{fig:resultadobounded2}
\end{figure}		
\begin{figure}[htb]
	\centering
	\centerline{\includegraphics[width=\linewidth]{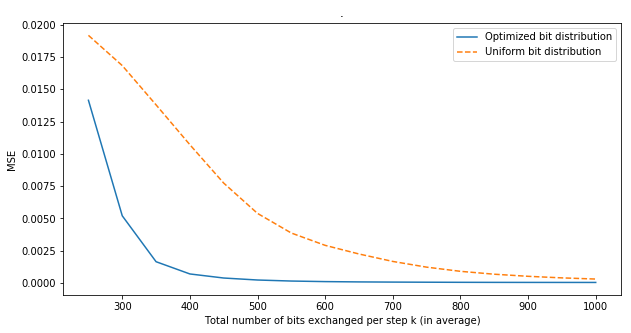}}
	\caption{Average MSE vs average number of bits for uniform and optimized bit allocation for $K=17$ and $\theta=0.001$ ($\kappa=0.2$, $ f_{in}[n]= a[n]^2+b[n]^2-1 $,  $ g(\lambda) = \frac{3}{3+5\lambda}   $).}
	\medskip
	\label{fig:resultadobounded3}
\end{figure}


It can be seen from Fig. \ref{fig:resultadobounded} that the optimization of the bit allocation proposed in this paper clearly improves the performance in terms of MSE if compared to a quantizer with the uniform bit distribution. A bigger gain appears at low bit rate where effective allocation is more important as resource are scarcer.

To evaluate the efficiency of the bounding scheme, we process the signal using the distributed algorithm proposed in \cite{7447171}, where the messages are not bounded. The settings are the same as the previous experiment. We obtain a MSE of 0.77 for 300 exchanged bits, a MSE of 0.55 for 400 exchanged bits and a MSE of 0.26 for 600 exchanged bits. These results show that, regardless of optimizing the bit distribution or not, our bounded range algorithm yields much lower MSE values compared to the baseline algorithm for the same bit rates.  This is due to the fact that, without bounding the transmitted messages, they grow substantially and in consequence, the quantization errors grow as well.

We can analyze the effects of the graph structure and the number of iterations on the performance of the optimization scheme. First, the same processing (with bounding scheme) is applied into a graph with a smaller $\theta$, which results in a bigger discrepancy between edges weights, that is, a less regular graph. The communication constraints are however unchanged, since the number of edges only depends on $\kappa$, which remains constant in our experiments. The results can be seen in Fig. \ref{fig:resultadobounded2}, where the difference between the uniform and the optimized bit distribution is slightly bigger than in the original experiments with a more regular graph (Fig. \ref{fig:resultadobounded}). It means that a more regular graph tends towards a more uniform bit distribution, which seems reasonable. Finally we look at the impact of the polynomial order $K$ on the performance. We run experiments similar to the previous ones, but with a bigger value of $K$. The difference between the uniform and the optimized bit allocations is shown in Fig. \ref{fig:resultadobounded3}. We see that the gain due to optimal allocation is bigger than in the previous experiments with the same network (Fig. \ref{fig:resultadobounded2}). When $k$ grows, the errors propagate more across iterations and the optimized bit allocation tries to compensate it by allocating more bits in the first iteration steps. This allocation improves the performance of the optimized scheme with respect to the uniform scheme, and this improvement effect is more visible for greater $K$, since the errors propagate for more iterations.

\subsubsection{Performance with different filters}
\begin{figure}[htb]
	\centering
	\centerline{\includegraphics[width=\linewidth]{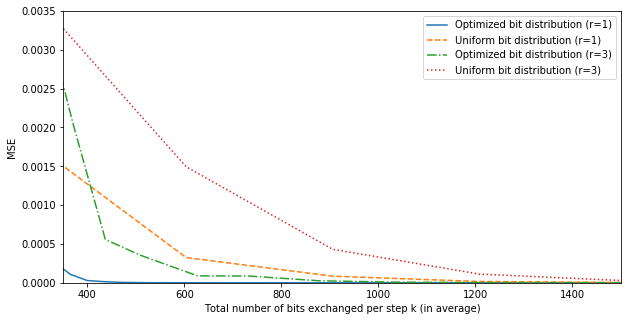}}
	\caption{Average MSE vs average number of bits for distributed Tikhonov regularization with optimal and uniform bit allocation for $r=1$ and $r=3$ ($\tau=10$, $\theta=0.001$, $\kappa = 0.2$, $K=9$, $ f_{in}[n]= a[n]^2+b[n]^2-1 $).}\medskip
	\label{tikonov}
\end{figure}

\begin{figure}[htb]
	\centering
	\centerline{\includegraphics[width=\linewidth]{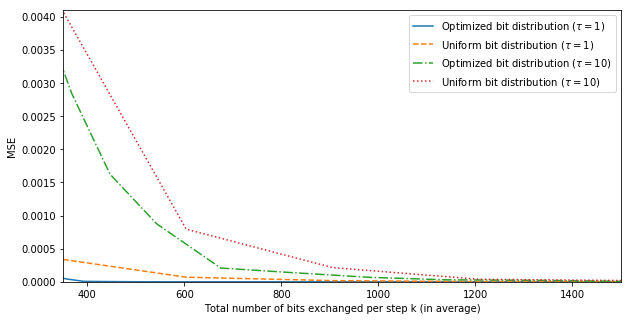}}
	\caption{Average MSE vs average number of bits for distributed smoothing  with optimal and uniform bit allocation for $\tau=1$ and $\tau=10$ ($\theta=0.001$, $\kappa = 0.2$, $K=9$, $ f_{in}[n]= a[n]^2+b[n]^2-1 $).}\medskip
	\label{heat}
\end{figure}

We now analyze how the performance of our algorithm changes with different filters implemented in the distributed processing task. Graphs are generated in the same way as in Subsection \ref{sub_perf_opt_quant} with $ \theta=0.001 $. The same input graph signal is now filtered by different operators with the polynomial approximation order fixed at $ K=9 $. Two different tasks are performed with varying filter parameters. First, we perform denoising with distributed Tikhonov regularization. Such a task can be implemented by applying the graph filter
\begin{equation}
g(\lambda)=\frac{\tau}{\tau+2\lambda^r}
\end{equation}
to the input signal. We fix the value of $\tau$ at 10, and we choose for $r$ the values of 1 or 3. The second task consists in distributed smoothing, which can be performed by applying the heat kernel
\begin{equation}
g(\lambda) = e^{-\tau \frac{\lambda}{\lambda_{\text{max}}} }
\end{equation}
to the input signal. We consider $\tau$ taking the values 1 or 10. The results of the filter processing using different quantization methods are plotted in Figures  \ref{tikonov} and \ref{heat}.

Notice that, for the Tikhonov regularization, the higher $ r $, the less smooth the filter. As for the heat kernel, the higher $  \tau $, the less smooth the filter. As we can see in Figures \ref{tikonov} and \ref{heat}, when the filter becomes smoother, the values of the MSE become smaller (both for optimized and uniform bit allocation). This happens because smooth filters tend to have smaller values of $  \alpha_k $, the filter coefficients derived from the Chebyshev polynomial approximation. Thus, the quantization errors are multiplied by smaller factors, resulting in smaller quantization error. 

\subsubsection{Performance with different input signals}

\begin{figure}[htb]
	\centering
	\centerline{\includegraphics[width=\linewidth]{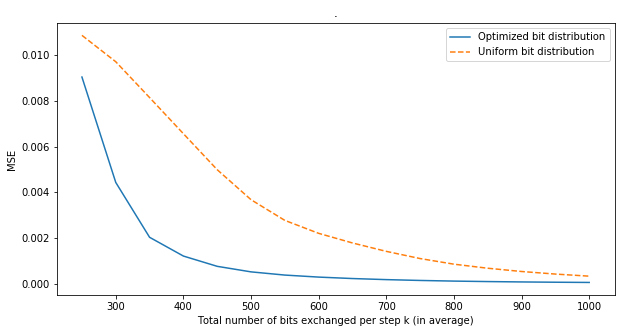}}
	\caption{Average MSE vs average number of bits for uniform and optimized bit allocation ($K=9$, $\theta=0.001$, $\kappa=0.2$, $ g(\lambda) = \frac{3}{3+5\lambda}    $), for random input signals.}\medskip
	\label{varios_fin}
\end{figure}

We now analyze the performance for different input graph signals. A graph is generated in the same way as in subsection \ref{sub_perf_opt_quant} with $ \theta=0.001 $, but now the input signal is a uniform random vector, whose values vary from 0 to 1. A Gaussian noise with zero-mean is added to the signal and the denoising is performed by distributively applying the same low-pass filter of Eq. (\ref{filtro}) to the signal, with its Chebyshev polynomial approximation of order $K=9$. The experiment is repeated 1000 times, with a different input signal at each time (while the other parameters remain fixed) and the average performance is computed and shown in Figure \ref{varios_fin}. We can see that the optimized bit allocation improves the filtering performance compared to the uniform bit distribution. 

\subsubsection{Performance on a real dataset}
\begin{figure}[htb]
	\centering
	\centerline{\includegraphics[width=\linewidth]{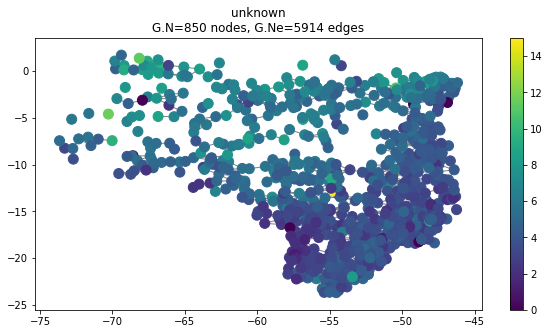}}
	\caption{Graph of the rain dataset. Each node represents the rain gauge stations. The colors represent the signal defined on the graph, that is, the daily average of rain in [mm/day] in the year 2010, at each station }\medskip
	\label{fig:RainG}
\end{figure}

We now consider a dataset containing the rain gauge time series components of a curated set of historical daily rainfall data for the Amazonian rainforest region in Brazil \cite{levy_2017}. The data is provided by the Brazillian water management agency Agencia Nacional de Aguas (ANA). Daily rainfall intensities in [mm/day] were recorded by 850 rain gauges spread in a large region spanning the southern Amazonian rainforest to the Cerrado biomes of Brazil from 1926 to 2013. We define a geographical graph, where the nodes of the graph consist of the rain gauge stations. The weight matrix is generated based on a thresholded Gaussian kernel function that takes into account the geographical distance between stations. The year of 2010 was chosen for containing the most complete data, and the average daily rain of that year was computed for each rain site to form the signal value at each node. This results in a graph signal of dimension 850. This graph can be visualized in Fig. \ref{fig:RainG}.

\begin{figure}[htb]
	\centering
	\centerline{\includegraphics[width=\linewidth]{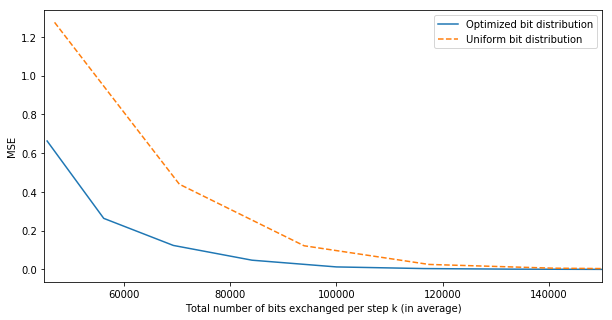}}
	\caption{MSE between filter outputs in perfect settings and in quantized settings vs average number of bits for the rain dataset ($K=9$, $\kappa=1.8$, $\theta=0.05$ and $ g(\lambda) = \frac{3}{3+5\lambda}    $). }\medskip
	\label{figRain_opt}
\end{figure}

A Gaussian noise with zero-mean is added to the rain signal and denoising is performed by distributively applying the same low-pass filter as in  Eq. (\ref{filtro})  (actually its Chebyshev polynomial approximation of degree $K$) to the signal and results are shown in Fig \ref{figRain_opt}. When the signal is  processed with the algorithm proposed in \cite{7447171}, where the messages are not bounded,  we obtain a MSE of 93.7 for 60000 exchanged bits and a MSE of 87.5 for 80000 exchanged bits. Comparing these results with Fig. \ref{figRain_opt}, we can see that our proposed modified algorithm yields much lower MSE values compared to the baseline algorithm. Also in Fig. \ref{figRain_opt} we see that the optimized bit allocation further improves the MSE in the bounded scheme, when compared to the uniform bit distribution.

The rain signal is smooth, and the filter is the same as the one used in Subsection \ref{sub_perf_opt_quant}, so the differences in amplitude observed between Fig. \ref{figRain_opt}  and the results in Subsection \ref{sub_perf_opt_quant} stem from the range of the input signal and the topology. The graph of the rain topology dataset has more nodes, which increases the number of computations necessary to filter the signal. With the increase in the number of computations, more errors are committed and accumulated, hence why the MSE values tend to be higher.

\subsection{Analysis of the bit allocation} \label{sec:bit}

\subsubsection{Evolution of $  F_k[n] $} \label{sec:fk}

\begin{figure}[htb]
	\centering
	\centerline{\includegraphics[width=\linewidth]{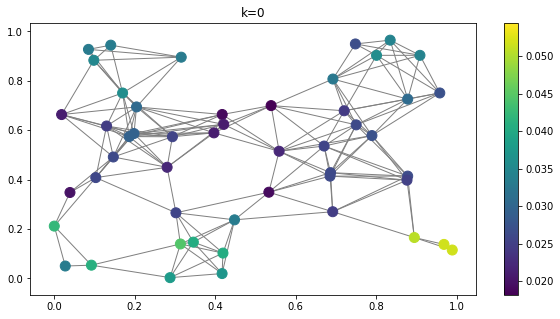}}
	\caption{ $  F_k[n] $  at step $ k=0 $. The colors represent the values of $  F_0[n] $ at different nodes ($K=9$, $\theta = 2$, $\kappa=0.2$, $ g(\lambda)=\frac{3}{3+5\lambda}    $, $ f_{in}[n]= a[n]^2+b[n]^2-1 $ ).}\medskip
	\label{F_N}
\end{figure}

Since $F_k[n]$ influences the optimal bit allocation $x[n,k]$, it is important to understand which factors influence it. A new graph is generated in the same way as in Subsection \ref{sub_perf_opt_quant}, also the same graph signal is considered to be filtered by the same filter of Eq. (\ref{filtro}) with an approximation of order $ K = 9 $ and edge weight scale factor $\theta=2$.  We compute $  F_k[n] $ with Eq. (\ref{eqt}). Now we observe the relationship of $  F_k[n] $ with $ n $ and $ k $. In order to analyze the relation with $ n $ first, we plot $  F_0[n] $ over the graph in Fig. \ref{F_N}, namely $  F_k[n] $ at step $ k=0 $. At this initial step we have the highest variance of $F_k[n]$, so we can visualize its relationship with $n$ better. The colors represent the values of $  F_0[n] $ at different nodes. It can be noted that groups of nodes that are more isolated tend to have higher values of $  F_k[n] $. As a result, a higher number of bits needs to be allocated to those nodes.  This means that the contribution to the global error of these nodes becomes higher. That happens because isolated nodes tend to amplify their errors, whereas more connected nodes tend to dissolve their errors into the network.

\begin{figure}[htb]
	\centering
	\centerline{\includegraphics[width=\linewidth]{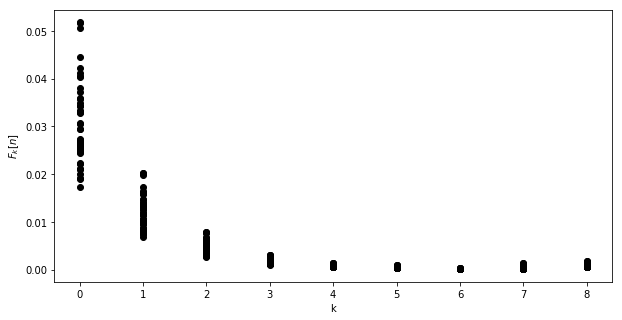}}
	\caption{ $ F_k[n]  $  versus $ k $ for all nodes  ($K=9$, $\theta = 2$, $\kappa=0.2$, $ g(\lambda)=\frac{3}{3+5\lambda}    $, $ f_{in}[n]= a[n]^2+b[n]^2-1 $ ).}\medskip
	\label{F_K}
\end{figure}

We now plot in Fig. \ref{F_K} the values of $ F_k[n] $  versus $ k $. For a given $k$, we show different values of $  F_k[n] $, one value for each $n$. We observe that the general trend is to $  F_k[n] $ decrease with $ k $, as expected since initial errors indeed propagate more than the later quantization errors as discusses earlier.

\subsubsection{Variance of $x[n,k]$}

We now have a deeper look at the actual optimal bit allocation in order to understand for which topologies the variance of $x[n,k]$ is high or low. A low variance means that the optimal solution approaches the one of a uniform solution. In this case, it might be simpler to perform a uniform bit allocation even with slightly worse MSE results.

In another set of experiments, 300 graphs generated from a weight matrix based on a thresholded Gaussian kernel function were generated with the edge weight scale factor $ \theta $ fixed at $ 2 $. The edge threshold value $ \kappa $ varies from $ 0.15 $ to $ 0.4 $, so that we can have graphs with different values of mean degree and variance. The unconnected graphs were removed from the experiment. The graphs were chosen with $ N=50 $ nodes. The same graph signal as in Subsection \ref{sub_perf_opt_quant} is generated and filtered with the same filter of Eq. (\ref{filtro}) with polynomial approximation order $  K=9 $.  We use Eq. (\ref{eqw}) for computing $x[n,k]$. 

\begin{figure}[htb]
	\centering
	\centerline{\includegraphics[width=\linewidth]{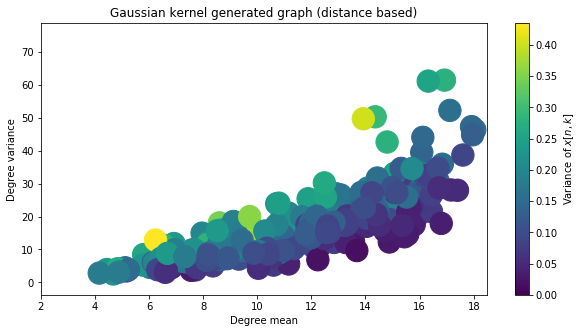}}
	\caption{Relationship between the degrees' variance and mean with the variance of $x[n,k]$ for graphs generated from a weight matrix based on a thresholded gaussian kernel function. The colors represent the variance of $x[n,k]$. Each dot is a different graph.}\medskip
	\label{gaussian1}
\end{figure}

In Fig. \ref{gaussian1}, we can see the relationship between the degrees' variance and mean with the variance of $x[n,k]$ with respect to $ n $ ($x[n,k]$ is averaged with respect to $ k $). Each dot corresponds to the experiment resulted with the use of a different graph. It can be noted from Figs. \ref{gaussian1}  that if the graph has low degree variance, that is, it is a more regular graph, it tends to have a more uniform bit distribution. At the same time, graphs with higher degree mean (that is, denser graphs) also tend to have more uniform bit distribution, which is expected since more connected graphs tend to have the quantization errors smoothed out more uniformly in the network. These facts mean that when the graph is both sparser and irregular it will require a more irregular bit distribution hence optimal bit allocation is important. 

\begin{figure}[htb]
	\centering
	\centerline{\includegraphics[width=\linewidth]{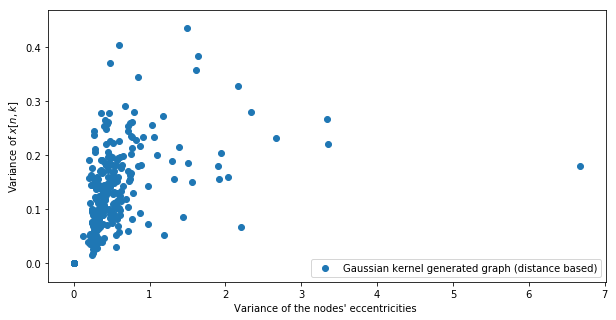}}
	\caption{Variance of  $x[n,k]$ versus variance of the nodes' eccentricities for Gaussian kernel generated graphs.}\medskip
	\label{_e_gaussian}
\end{figure}

In Fig. \ref{_e_gaussian}, we can further see the relationship with the variance of the nodes' eccentricities. The eccentricity of a graph node $n$ is the maximum distance between $n$ and any other node of the graph.  In  Fig. \ref{_e_gaussian} we can also notice the tendency that graphs with more irregular eccentricities require more irregular bit distribution, because more eccentric nodes require more bits than the less eccentric ones, as discussed previously. Which means that, for graphs with more irregular eccentricities, the benefit of solving the optimal bit allocation problem is higher.


\section{Conclusion}

In this paper, we have shown how the quantization error in distributed graph signal processing tasks can be minimized by bounding the transmitted messages and by optimizing the bit allocation in the network. Our method is important in cases where we have a bit budget constraint. The optimal bit allocation varies with the network nodes and the steps of the iterative filtering process; more specifically, nodes at the border of the network and the initial steps of the iterative algorithm tend to require more bits. Its variance varies with the topology, where irregular and sparse topologies lead to high variance of the optimal bit allocation. Experimental results show that our distributed processing algorithm substantially decreases the quantization error with regard to previous solutions and to a uniform bit allocation scheme. Future work will investigate non-uniform quantization for each message (varying quantization step size), which will improve the performance in low-rate regime.

\ifCLASSOPTIONcaptionsoff
  \newpage
\fi

\bibliographystyle{IEEEbib}
\bibliography{Referencias}

\begin{thebibliography}{10}

\bibitem{ortega2017graph}
Antonio Ortega, Pascal Frossard, Jelena Kova{\v{c}}evi{\'c}, Jos{\'e}~MF Moura,
  and Pierre Vandergheynst,
\newblock ``Graph signal processing,''
\newblock {\em arXiv preprint arXiv:1712.00468}, 2017.

\bibitem{6494675}
David~I Shuman, Sunil~K Narang, Pascal Frossard, Antonio Ortega, and Pierre
  Vandergheynst,
\newblock ``The emerging field of signal processing on graphs: Extending
  high-dimensional data analysis to networks and other irregular domains,''
\newblock {\em IEEE signal processing magazine}, vol. 30, no. 3, pp. 83--98,
  2013.

\bibitem{DBLP:journals/corr/abs-1111-5239}
David~I Shuman, Pierre Vandergheynst, Daniel Kressner, and Pascal Frossard,
\newblock ``Distributed signal processing via chebyshev polynomial
  approximation,''
\newblock {\em IEEE Transactions on Signal and Information Processing over
  Networks}, vol. 4, no. 4, pp. 736--751, 2018.

\bibitem{sandryhaila2014finite}
Aliaksei Sandryhaila, Soummya Kar, and Jos{\'e}~MF Moura,
\newblock ``Finite-time distributed consensus through graph filters,''
\newblock in {\em IEEE ICASSP}, 2014, pp. 1080--1084.

\bibitem{safavi2015revisiting}
Sam Safavi and Usman~A Khan,
\newblock ``Revisiting finite-time distributed algorithms via successive
  nulling of eigenvalues,''
\newblock {\em IEEE Signal Processing Letters}, vol. 22, no. 1, pp. 54--57,
  2015.

\bibitem{isufi2017autoregressive}
Elvin Isufi, Andreas Loukas, Andrea Simonetto, and Geert Leus,
\newblock ``Autoregressive moving average graph filtering,''
\newblock {\em IEEE Transactions on Signal Processing}, vol. 65, no. 2, pp.
  274--288, 2017.

\bibitem{segarra2017optimal}
Santiago Segarra, Antonio~G Marques, and Alejandro Ribeiro,
\newblock ``Optimal graph-filter design and applications to distributed linear
  network operators,''
\newblock {\em IEEE Transactions on Signal Processing}, vol. 65, no. 15, pp.
  4117--4131, 2017.

\bibitem{shi2015infinite}
Xuesong Shi, Hui Feng, Muyuan Zhai, Tao Yang, and Bo~Hu,
\newblock ``Infinite impulse response graph filters in wireless sensor
  networks,''
\newblock {\em IEEE Signal Processing Letters}, vol. 22, no. 8, pp. 1113--1117,
  2015.

\bibitem{8682784}
Isabela~CM Nobre and Pascal Frossard,
\newblock ``Optimized quantization in distributed graph signal processing,''
\newblock in {\em ICASSP 2019-2019 IEEE International Conference on Acoustics,
  Speech and Signal Processing (ICASSP)}. IEEE, 2019, pp. 5376--5380.

\bibitem{sandryhaila2013discrete}
Aliaksei Sandryhaila and Jos{\'e}~MF Moura,
\newblock ``Discrete signal processing on graphs,''
\newblock {\em IEEE Transactions on Signal Processing}, vol. 61, no. 7, pp.
  1644--1656, 2013.

\bibitem{liu2017filter}
Jiani Liu, Elvin Isufi, and Geert Leus,
\newblock ``Filter design for autoregressive moving average graph filters,''
\newblock {\em arXiv preprint arXiv:1711.09086}, 2017.

\bibitem{chamon2017finite}
Luiz~FO Chamon and Alejandro Ribeiro,
\newblock ``Finite-precision effects on graph filters,''
\newblock in {\em IEEE GlobalSIP}, 2017, pp. 603--607.

\bibitem{zhu2016quantized}
Shengyu Zhu, Mingyi Hong, and Biao Chen,
\newblock ``Quantized consensus admm for multi-agent distributed
  optimization,''
\newblock in {\em IEEE ICASSP}, 2016, pp. 4134--4138.

\bibitem{zhu2016quantized2}
Shengyu Zhu and Biao Chen,
\newblock ``Quantized consensus by the admm: probabilistic versus deterministic
  quantizers,''
\newblock {\em IEEE Transactions on Signal Processing}, vol. 64, no. 7, pp.
  1700--1713, 2016.

\bibitem{li2017event}
Huaqing Li, Shuai Liu, Yeng~Chai Soh, and Lihua Xie,
\newblock ``Event-triggered communication and data rate constraint for
  distributed optimization of multiagent systems,''
\newblock {\em IEEE Transactions on Systems, Man, and Cybernetics: Systems}, ,
  no. 99, pp. 1--12, 2017.

\bibitem{li2017distributed}
Jueyou Li, Guo Chen, Zhiyou Wu, and Xing He,
\newblock ``Distributed subgradient method for multi-agent optimization with
  quantized communication,''
\newblock {\em Mathematical Methods in the Applied Sciences}, vol. 40, no. 4,
  pp. 1201--1213, 2017.

\bibitem{thanou2013distributed}
Dorina Thanou, Effrosyni Kokiopoulou, Ye~Pu, and Pascal Frossard,
\newblock ``Distributed average consensus with quantization refinement,''
\newblock {\em IEEE Transactions on Signal Processing}, vol. 61, no. 1, pp.
  194--205, 2013.

\bibitem{7447171}
Dorina Thanou and Pascal Frossard,
\newblock ``Learning of robust spectral graph dictionaries for distributed
  processing,''
\newblock {\em EURASIP Journal on Advances in Signal Processing}, vol. 2018,
  no. 1, pp. 67, 2018.

\bibitem{meyer2000matrix}
Carl~D Meyer,
\newblock {\em Matrix analysis and applied linear algebra}, vol.~71,
\newblock Siam, 2000.

\bibitem{492748}
Bernard Widrow, Istvan Kollar, and Ming-Chang Liu,
\newblock ``Statistical theory of quantization,''
\newblock {\em IEEE Transactions on instrumentation and measurement}, vol. 45,
  no. 2, pp. 353--361, 1996.

\bibitem{1166538}
Lin Xiao, Mikael Johansson, Haitham Hindi, Stephen Boyd, and Andrea Goldsmith,
\newblock ``Joint optimization of communication rates and linear systems,''
\newblock {\em IEEE Transactions on Automatic Control}, vol. 48, no. 1, pp.
  148--153, 2003.

\bibitem{boyd2004convex}
Stephen Boyd and Lieven Vandenberghe,
\newblock {\em Convex optimization},
\newblock Cambridge university press, 2004.

\bibitem{1055533}
Adrian Segall,
\newblock ``Bit allocation and encoding for vector sources,''
\newblock {\em IEEE transactions on Information Theory}, vol. 22, no. 2, pp.
  162--169, 1976.

\bibitem{levy_2017}
Morgan Levy,
\newblock ``Curated rain and flow data for the brazilian rainforest-savann
  transition zone,''
\newblock {\em HydroShare}, Mar 2017.

\end{thebibliography}

\appendices

\section{Bit allocation in the network }

\begin{figure}[htb]
	\centering
	\centerline{\includegraphics[width=0.8\linewidth]{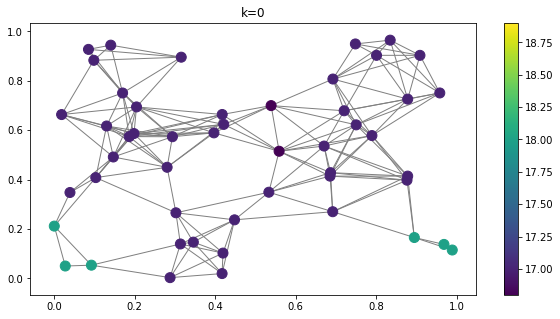}}
	\caption{$  x[n,k] $  at step $ k=0 $. The colors represent the values of $  x[n,0] $ at different nodes ($K=9$, $\theta = 2$, $\kappa=0.2$, $ g(\lambda)=\frac{3}{3+5\lambda}    $, $ f_{in}[n]= a[n]^2+b[n]^2-1 $ ).}\medskip
	\label{X_otimo}
\end{figure}

We now observe the relationship between $  F_k[n] $ and $x[n,k]$. In the same experiment of Subsection \ref{sec:fk}, we show in Fig. \ref{X_otimo} the values of $x[n,k]$ plotted as the colors of the graph. Comparing Figs. \ref{F_N} and \ref{X_otimo}, we notice that, when the values of $x[n,k]$ are high,  they are also high in $  F_k[n] $; but the opposite does not necessarily hold true. This happens because the values of $  F_k[n] $  have the tendency of being higher on the more isolated nodes, but the values of $x[n,k]$ take into consideration communication costs too. Thus they tend to be higher on those nodes that are not only at the border of the graph but have also small degrees. 

\section{Error Propagation}

\begin{figure}[htb]
	\centering
	\centerline{\includegraphics[width=0.8\linewidth]{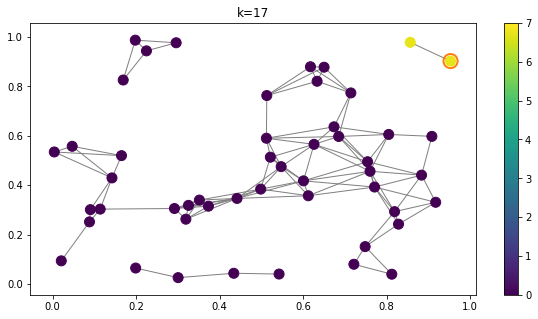}}
	\caption{Graph built with $\kappa=0.18$. The colors represent the absolute difference between $z_{17}[n]$ in an unquantized processing (true value) and its value in the experiment where only  $z_0[n]$ is quantized but all the other values (different $ k $s or different $ n $'s) are not. The chosen node $n$  is highlighted within an orange circle. The global MSE is $ 6.97e-07 $.}\medskip
	\label{firststep_018}
\end{figure}

\begin{figure}[htb]
	\centering
	\centerline{\includegraphics[width=0.8\linewidth]{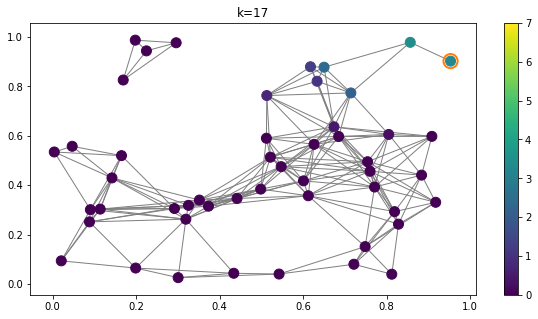}}
	\caption{Graph built with $\kappa=0.25$. The colors represent the absolute difference between $z_{17}[n]$ in an unquantized processing (true value) and its value in the experiment where only  $z_0[n]$ is quantized but all the other values (different $ k $s or different $ n $'s) are not. The chosen node $n$  is highlighted within an orange circle. The  global MSE is $ 1.10e-07$.}\medskip
	\label{firststep_025}
\end{figure}

\begin{figure}[htb]
	\centering
	\centerline{\includegraphics[width=0.8\linewidth]{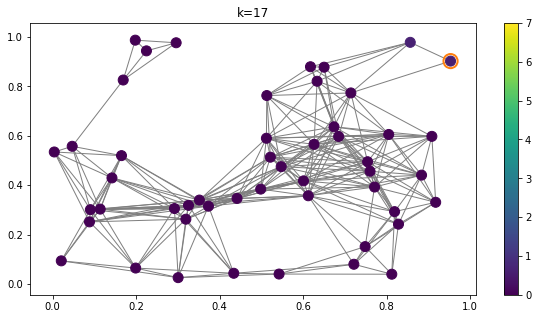}}
	\caption{Graph built with $\kappa=0.3$. The colors represent the absolute difference between $z_{17}[n]$ in an unquantized processing (true value) and its value in the experiment where only  $z_0[n]$ is quantized but all the other values (different $ k $s or different $ n $'s) are not. The chosen node $n$  is highlighted within an orange circle. The global MSE is $  0.47e-7  $.}\medskip
	\label{firststep_030}
\end{figure}

We do an analysis of the error propagation for illustration purpose. Three graphs are generated in the same way as in subsection \ref{sub_perf_opt_quant}. The same input graph signal is filtered by distributively applying the same low-pass filter of Eq. (\ref{filtro}) to the signal, with its Chebyshev polynomial approximation of order $ K = 17$. We fix $\theta=2$ and vary $\kappa$. When $\kappa$ is small, the resulting graph is sparse. Thus, the three generated graphs have the same set of nodes but have different sparsity values, according to the chosen value of $\kappa$.  Figs. \ref{firststep_018}, \ref{firststep_025} and \ref{firststep_030} show the three graphs for the values of $ \kappa $ = 0.18, 0.25 and 0.3 respectively.

We are interested in observing how the individual error $\epsilon_0[n]$, at a specific value of $n$ and generated in the first iteration step $k=0$, is propagated through the network.  To achieve this, we perform an experiment where all the other nodes send their messages without quantization  (that is, as the perfect representation of their true value) and node $n$ only quantizes at step $k=0$, while for the next values of $k$ it sends the messages unquantized. By doing this we can assure that all observed errors in the distributed signal processing stem from  $\epsilon_0[n]$ and its propagation. The bounded scheme proposed in this paper is used for all cases.

The chosen node $n$  is highlighted within an orange circle. The colors represent the absolute difference between $z_{17}[n]$ in an unquantized processing (true value) and its value in the experiment mentioned where only  $z_0[n]$ is quantized but all the other values (different $ k $'s or different $ n $'s) are not. We choose to represent this difference at k=17 since it is the filtering final step and we can thus see the accumulated propagation of $\epsilon_0[n]$.  The difference is plotted in log scale. 

We can see in Fig. \ref{firststep_018}  that the node $ n $ is only connected to another node and disconnected from the rest. Thus, $\epsilon_0[n]$ does not propagate to most of the nodes resulting in these unconnected nodes having no errors on $z_{17}[n]$. On the other hand, node $ n $ and its neighbor present very high error, which suggests that the error was amplified among them. When we compare to Figs. \ref{firststep_025} and \ref{firststep_030}, where the node $ n $ is connected to the rest of the network, $\epsilon_0[n]$ propagates to the other nodes and thus the value of the error at $n$ drops considerably. When we consider the last Figure in particular, we notice that there is barely any error in any node. The global MSE for each case is $ 6.97e-07 $ for $\kappa=0.18$, $ 1.10e-07 $ for $\kappa=0.25$ and $ 0.47e-7 $ for $\kappa =0.3$, respectively. Notice that Fig. \ref{firststep_018} has the highest eccentricity for node $n$, which seems to suggest that nodes with high eccentricity, that is, more isolated nodes, tend to have their errors amplified, whereas the ones that are more central tend to have their errors dissolved into the network. This is coherent to what we previously observed in Section \ref{sec:bit}.


\section{Variance of  the bit allocation for binomial distribution}

Another 300 graphs were generated considering node degrees that follow a binomial distribution. This distribution was chosen was because the mean and variance of this distribution can be chosen independently (as long as the variance is below the mean). The graphs were chosen with $ N=50 $ nodes. The same graph signal as in Subsection \ref{sub_perf_opt_quant} is generated and filtered with the same filter of Eq. (\ref{filtro}) with polynomial approximation order $  K=9 $.  We use Eq. (\ref{eqw}) for computing $x[n,k]$. 

\begin{figure}[htb]
	\centering
	\centerline{\includegraphics[width=0.8\linewidth]{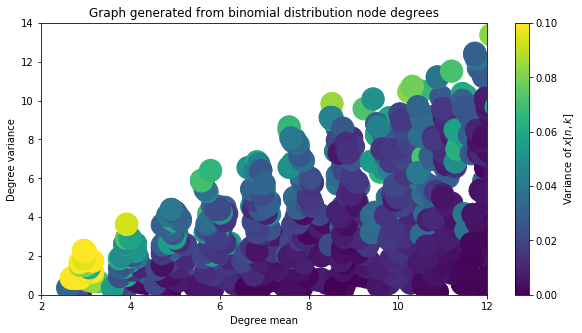}}
	\caption{Relationship between the degrees' variance and mean with the variance of $x[n,k]$ for graphs generated with node degrees following a binomial distribution. The colors represent the variance of $x[n,k]$ along $n$ and $k$. Each color dot is a different graph.}\medskip
	\label{binomial1}
\end{figure}

In Fig. \ref{binomial1}, we can see the relationship between the degrees' variance and mean with the variance of $x[n,k]$ with respect to $ n $ ($x[n,k]$ is averaged with respect to $ k $). Each dot corresponds to the experiment resulted with the use of a different graph. We can notice that the results of Fig. \ref{binomial1} are similar to those of Fig. \ref{gaussian1}, which means our observations are consistent for different types of graphs. 

\begin{figure}[htb]
	\centering
	\centerline{\includegraphics[width=0.8\linewidth]{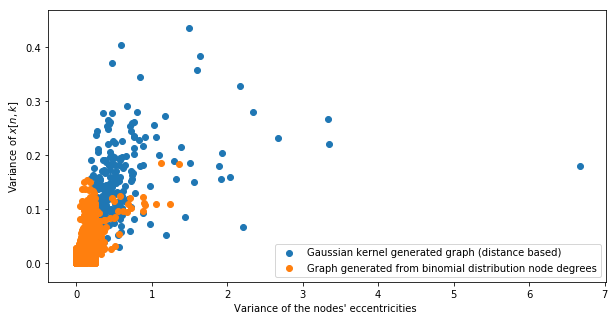}}
	\caption{Variance of  $x[n,k]$ versus variance of the nodes' eccentricities for two different type of graphs. In orange, we have graphs generated from binomial distribution node degrees and in blue we have Gaussian kernel generated graphs.}\medskip
	\label{binomial_e_gaussian}
\end{figure}

In Fig. \ref{binomial_e_gaussian}, we can further see the relationship with the variance of the nodes' eccentricities. In orange, we have graphs generated from binomial distribution node degrees and in blue we have Gaussian kernel generated graphs. Here, the same tendency is again noticed for both types of graphs.


\end{document}